\documentclass[10pt,aps,prx,twocolumn,superscriptaddress,floatfix,showpacs,longbibliography]{revtex4-1}
\usepackage[utf8]{inputenc}
\usepackage[T1]{fontenc}
\usepackage{graphicx}
\usepackage{tabularx}
\usepackage[usenames,dvipsnames,table]{xcolor}
\usepackage[version=3]{mhchem}
\usepackage{braket}
\usepackage{upgreek}
\usepackage{hyperref}
\usepackage{amsmath, amssymb}
\usepackage{bbold}
\usepackage[normalem]{ulem}
\usepackage{soul}

\definecolor{mark}{rgb}{0.85, 0.9, 1}

\hypersetup{hidelinks}
\sethlcolor{mark}

\renewcommand{\vec}[1]{\mathbf{#1}}

\begin{document}

\title{Emergence of global receptive fields capturing multipartite quantum correlations}

\author{Oleg M. Sotnikov}
\affiliation{Theoretical Physics and Applied Mathematics Department, Ural Federal University, Ekaterinburg 620002, Russia}
\affiliation{Russian Quantum Center, Skolkovo, Moscow 121205, Russia}
\author{Ilia A. Iakovlev}
\affiliation{Theoretical Physics and Applied Mathematics Department, Ural Federal University, Ekaterinburg 620002, Russia}
\affiliation{Russian Quantum Center, Skolkovo, Moscow 121205, Russia}
\author{Evgeniy O. Kiktenko}
\affiliation{Russian Quantum Center, Skolkovo, Moscow 121205, Russia}
\affiliation{National University of Science and Technology ``MISIS'', Moscow 119049, Russia}
\author{Mikhail~I.~Katsnelson}
\affiliation{Institute for Molecules and Materials, Radboud University, Heyendaalseweg 135, 6525AJ, Nijmegen, Netherlands}
\author{Aleksey K. Fedorov}
\affiliation{Russian Quantum Center, Skolkovo, Moscow 121205, Russia}
\affiliation{National University of Science and Technology ``MISIS'', Moscow 119049, Russia}
\author{Vladimir V. Mazurenko}
\affiliation{Theoretical Physics and Applied Mathematics Department, Ural Federal University, Ekaterinburg 620002, Russia}
\affiliation{Russian Quantum Center, Skolkovo, Moscow 121205, Russia}

\date{\today}

\begin{abstract}
In quantum physics, even simple data with a well-defined structure at the wave function level can be characterized by extremely complex correlations between its constituent elements. The inherent non-locality of the quantum correlations generally prevents one from providing their simple and transparent interpretation, which also remains a challenging problem for advanced classical techniques that approximate quantum states with neural networks. Here we show that monitoring the neural network weight space while learning quantum statistics from measurements allows to develop physical intuition about complex multipartite patterns and thus helps to construct more effective classical representations of the wave functions. Particularly, we observe the formation of distinct global convolutional structures, receptive fields in the hidden layer of the Restricted Boltzmann Machine (RBM) within the neural quantum tomography of the highly-entangled Dicke states. On this basis we propose an exact two-parameter classical representation not only for a specific quantum wave function, but for the whole family of the $N$-qubit Dicke states of different entanglement. Our findings suggest a fresh look at constructing convolutional neural networks for processing data with non-local patterns and pave the way for developing exact learning-based representations of entangled quantum states. 
\end{abstract}

\maketitle

\section{Introduction}
Machine leaning is among the most promising modern technologies, which has a vast impact on all areas of activities. Recognizing and classifying objects, translating languages, generating texts~\cite{chatGPT} and videos, creating images~\cite{kandinsky}, playing games~\cite{Atari, Go, Poker}, predicting human lives~\cite{predict_live} are only some examples of the tasks that can be solved with modern artificial neural networks. In scientific domain, neural networks accelerate search for novel materials~\cite{Deepmind_materials}, recognize different phases of matter~\cite{Melko}, facilitate simulations of complex quantum systems~\cite{Carleo, Imada} with the number of particles inaccessible to standard methods such as exact diagonalization, help to discover quantum-error-correction strategies against noise~\cite{Florian} and solve many other problems. It also suggests a new scientific paradigm which can be applied far beyond ``exact sciences'', for example, in the evolutionary biology~\cite{biology}.

Such a success of neural networks in different areas is mainly related to the possibility of capturing spatial, temporal and other correlations in input data. In this respect the concept of the receptive fields (RF) first introduced in neuroscience~\cite{place_cell1, place_cell2, place_cell3} for describing spatial behavior of animals holds significant importance in developing convolutional neural networks (CNN) \cite{CNN-LeCun, ConvolutionalRBM}. RF assumes that the input image contains highly correlated repeating local features. To capture them each unit from the first hidden layer of CNN can interact with only a portion of the input neurons (sparse connectivity) and such hidden neurons related to different parts of the image are then combined in a feature map where they share the same weights. This makes CNN invariant to shift transformations (translations) and provides its success in solving different tasks \cite{CNN_example1,CNN_example2, CNN_example3}. Remarkably, it has been found that CNN can be used in quantum physics to approximate quantum wave functions defined on regular lattices \cite{Choo}. Here, introducing rotations, mirror symmetries and other transformations \cite{GCNN1} into consideration positively affects the solution of complex problems in condensed matter physics~\cite{GCNN2}. 
 
The use of the RF concept that the input is characterized by some topology is not limited to CNN, it can be realized for other neural network architectures. For instance, the account of the regular lattice symmetries when simulating quantum spin Hamiltonians with Restricted Boltzmann Machines allows to reduce the total number of variational parameters and considerably improve the accuracy of the modeling \cite{Carleo}. Interestingly, the account of symmetries leads to formation of the receptive fields capturing quantum correlations in one- and two-dimensional magnetic systems. 

\begin{figure}[!t]
    \includegraphics[width=0.48\textwidth]{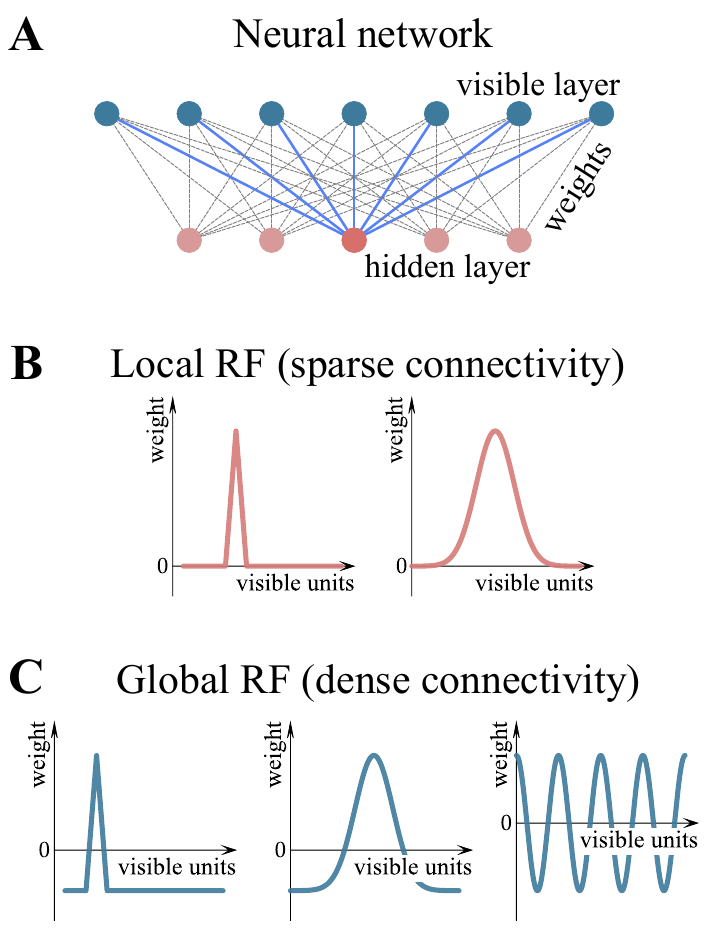}
	\caption{\label{intro} (A) An example of  a neural network fragment with highlighted couplings between visible neurons and a hidden neuron that can host a receptive field capturing correlations in a system. (B)  Weight profiles of the local receptive fields in the space of a hidden neuron obtained in previous works. (C) Examples of the weight-space profiles of the global receptive fields introduced in this work for representing 
 quantum states.}
\end{figure}

Another level of modeling RFs suggests detecting their emergence  when the neural network learns from scratch without imposing symmetries on the weights. As the bright example of this one can consider the RBM trained with Ising model configurations in Ref.~\cite{RF_Class_Ising} to reproduce the corresponding probabilities. Importantly, the shape of the resulting receptive fields was shown to reflect the magnetic correlations in the system in question. In general case preparing of the input data can play a decisive role in observing RFs. For instance, 
in Ref.~\cite{Conv_Dense} analyzing the weight space of a simple dense neural network trained by using the data with non-Gaussian high-order correlated structures, the authors have revealed and characterized receptive fields that resemble the filters of the convolutional neural networks. In quantum domain, the examples of the receptive field emergence have been found when solving the ground state problem for quantum XXZ model Hamiltonian with RBM representation of the wave function~\cite{Clark2}. Here the receptive fields can be classified as local, since each hidden unit has non-zero interaction with only one visible neuron. Similar local RF were analytically derived within the neural quantum state tomography~\cite{NNtomography} for some notable wave functions (see also Refs.~\cite{Tiunov2020,Fedotova2023}).  Typically, the size of the receptive fields spontaneously developed in hidden layers of neural networks is much smaller than the working space monitored by visible neurons. This reflects the finite length of the correlations in the input data and makes the situation beyond these limits practically unexplored.

In our paper, we introduce {\it the concept of a global receptive field, which describes the dense structured connectivity of a given hidden neuron with all visible units (Fig.\ref{intro}). Unlike the case of local RF described above, the corresponding functional dependence of weights has non-zero values located independently of the positions of connected elements}. We demonstrate that such RFs can be developed when the RBM neural network is trained to reproduce the state probabilities of a quantum system defined on the complete graph and characterized by multipartite correlations. For that, by using RBM we perform a neural quantum state tomography of the Dicke states \cite{Dicke_model} of different complexity. A remarkable property of thus found global RFs is that they emerge as the entanglement of the target quantum state increases. The coupling structure of our RFs differs from those reported in the previous works (Fig.\ref{intro} B). For instance, in the case of the Dicke states each hidden unit has a maximal positive coupling with a single visible neuron and nearly constant negative coupling with the rest ones (Fig.\ref{intro} left). At the same time, for other types of translationally invariant quantum states, global receptive fields with Gaussian- or cosine-like patterns in the couplings between hidden and visible neurons can be also found (Fig.\ref{intro} C center and right). It is important to note that our consideration is limited to real-valued wave functions with positive amplitudes. 

Undoubtedly, the crucial impact of learning RFs is the possibility to create and explore advanced neural network architectures, which allows reducing the total number of the network parameters and obtain more accurate solutions of a problem, as previously demonstrated in the case of the CNN. In our study we show that the learned global RFs can be used as filters for the restricted Boltzmann machine in the task of reconstructing quantum states of different entanglement. In this case a single compact RBM with the same number of the hidden and visible units allows exactly reconstructing the whole family of Dicke states from simple to complex ones.  

\section{Results}

\begin{figure*}[!t]
    \includegraphics[width=0.99\textwidth]{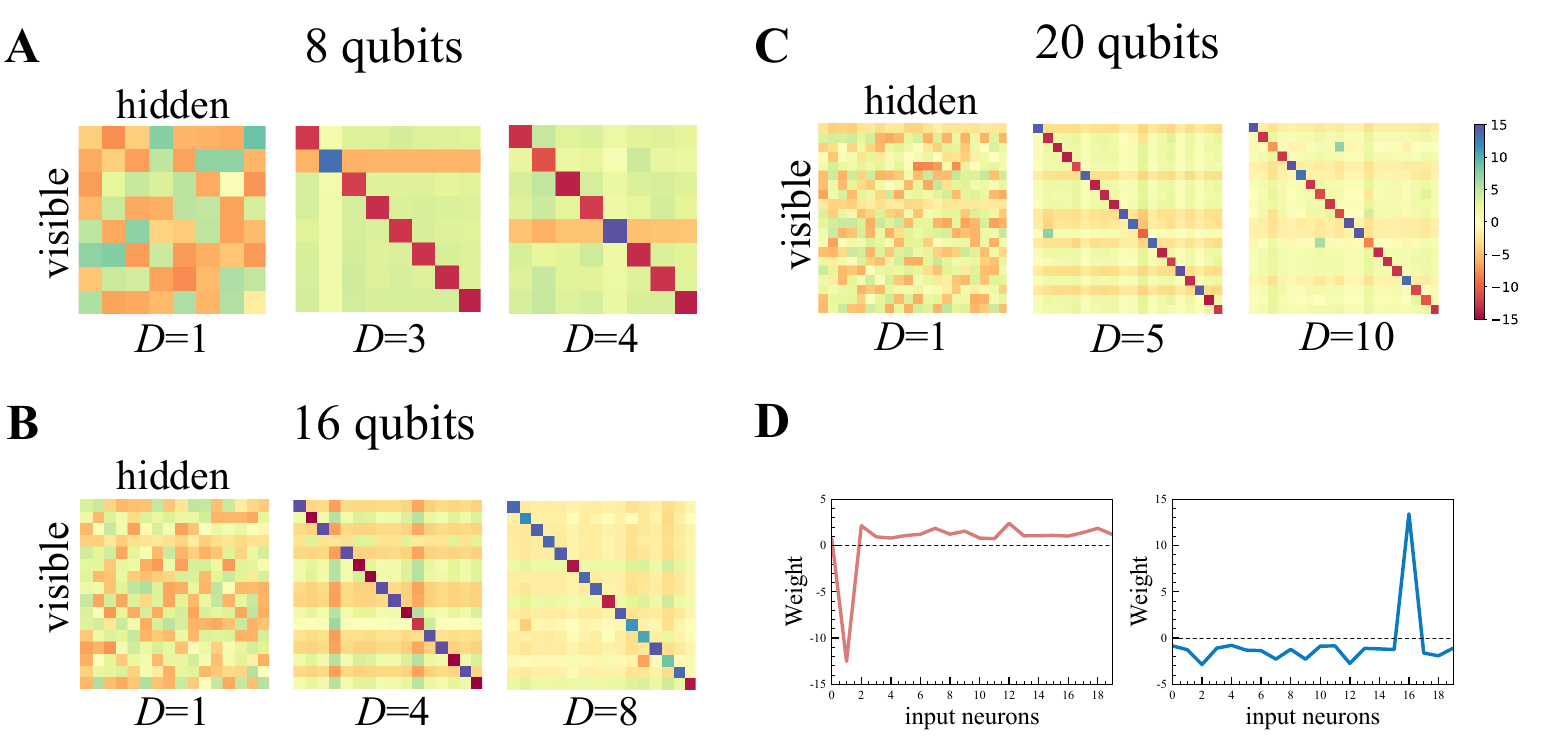}
	\caption{\label{16_tomography} Emergence of receptive fields in the weight space of the trained RBMs. (A-C) The neural quantum state tomography results obtained for the 8-, 16- and 20-qubit Dicke wave functions with different $D$ index. In all the cases the number of visible and hidden neurons is the same. The overlaps of the constructed RBMs with exact wave functions are 0.997 (A, $D=1$), 0.935 (A, $D=3$), 0.928 (A, $D=4$), 0.994 (B, $D=1$),  0.88 (B, $D=4$), 0.709 (B, $D=8$), 0.991 (C, $D=1$),  0.728 (C, $D=5$) and 0.605 (C, $D=10$). (D) Examples of the connectivity of two hidden neurons taken from the 20-qubit calculations with $D$=10. For all the presented examples the neural quantum state tomography procedure was performed with 10000 bitstring samples in the training set.}
\end{figure*}

\subsection{Learning correlations with global receptive fields}
We start with discussing the results of the neural quantum state tomography for the Dicke states,
\begin{eqnarray}
\Ket{\Psi^D_N} = \frac{1}{\sqrt{C^N_D}} \sum_{j} \mathcal {P}_{j}(\Ket{0}^{\otimes N-D} \otimes \Ket{1}^{\otimes D}),
\label{Dicke_wf}
\end{eqnarray}
where $N$ is the number of qubits, $D$ is the parameter that controls the number of ``1''s in each basis function contributing to the particular quantum state, $C_D^N=N!/[D!(N-D)!]$, and the sum goes over all possible permutations of qubits, denoted with $\mathcal{P}_j$.  Our main aim is to show that the receptive fields capturing non-local correlations can be discovered within the training without imposing symmetries. In the tomography approach the target wave function is to be approximated with positive neural quantum state
\begin{eqnarray}
  \label{eq:nqs}
\Ket{\Psi_{\boldsymbol{\lambda}}} = \sum_{{\bf v}} \Psi_{\boldsymbol{\lambda}} ({\bf v}) \Ket{{\bf v}}. 
\end{eqnarray}
Here, the amplitude $\Psi_{\boldsymbol{\lambda}}$ of a given basis state $\Ket{{\bf v}}$ is estimated with a neural network that is characterized by the set of weights, $\boldsymbol{\lambda}$ and takes the bitstring ${\bf v} = \{v_1,...v_N\}$ of the binary variables as the input. Generally, approximating quantum states can be fulfilled with using neural networks of different architectures including restricted Boltzmann machine \cite{Carleo, Imada, NNtomography, RF_TFI}, convolution \cite{Choo}, feed forward \cite{FFNN}, recurrent \cite{recurrent1, recurrent2} and transformer neural networks \cite{transformer1, transformer2}. 

Following the previous works devoted to neural quantum state tomography~\cite{NNtomography} of the Dicke wave functions, we employ RBM neural network as realized in the QuCumber package \cite{qucumber}. In this case the basis states are characterized by the amplitudes $\Psi_{\boldsymbol{\lambda}}({\bf v}) = \sqrt{p_{\boldsymbol{\lambda}}({\bf v})}$ and the corresponding probabilities  
\begin{eqnarray}
\label{eq:rbm}
p_{\boldsymbol{\lambda}}({\bf v}) = Z^{-1}_{\boldsymbol{\lambda}} e^{\sum_{i = 1}^{N} a_i v_i} \prod_{j=1}^{M} (1 + e^{b_{j} + \sum_{i=1}^{N} W_{ij} v_{i}}),
\end{eqnarray}
where $N$ ($M$) is the number of the visible (hidden) units, $\boldsymbol{\lambda} = \{ {\bf a}, {\bf b}, {\bf W} \}$ denotes the parameters of the restricted Boltzmann machine and the partition function $Z_{\boldsymbol{\lambda}}$ ensures the normalization of $p_{\boldsymbol{\lambda}}$. $W_{ij}$ is the weight matrix between visible and hidden units and $a_i$ ($b_j$) is the bias of the $i$th visible ($j$th hidden) neuron. The goal of the tomographic procedure is to optimize the network weights in such a way that the approximate wave function, Eq.\ref{eq:nqs} has the largest overlap with the target wave function that is one of the Dicke states, Eq.\ref{Dicke_wf}. For these purposes, one uses the Kullback-Leibler divergence~\cite{Amari} as the loss function. Importantly, the information about the target wave function is only represented in the form of finite sets of the bitstrings obtained from projective measurements. 
The comprehensive description of the neural quantum state tomography procedure we use can be found in Refs.~\cite{qucumber, Torlai_thesis, NNtomography}. 

\begin{figure*}[t]
	\includegraphics[width=\linewidth]{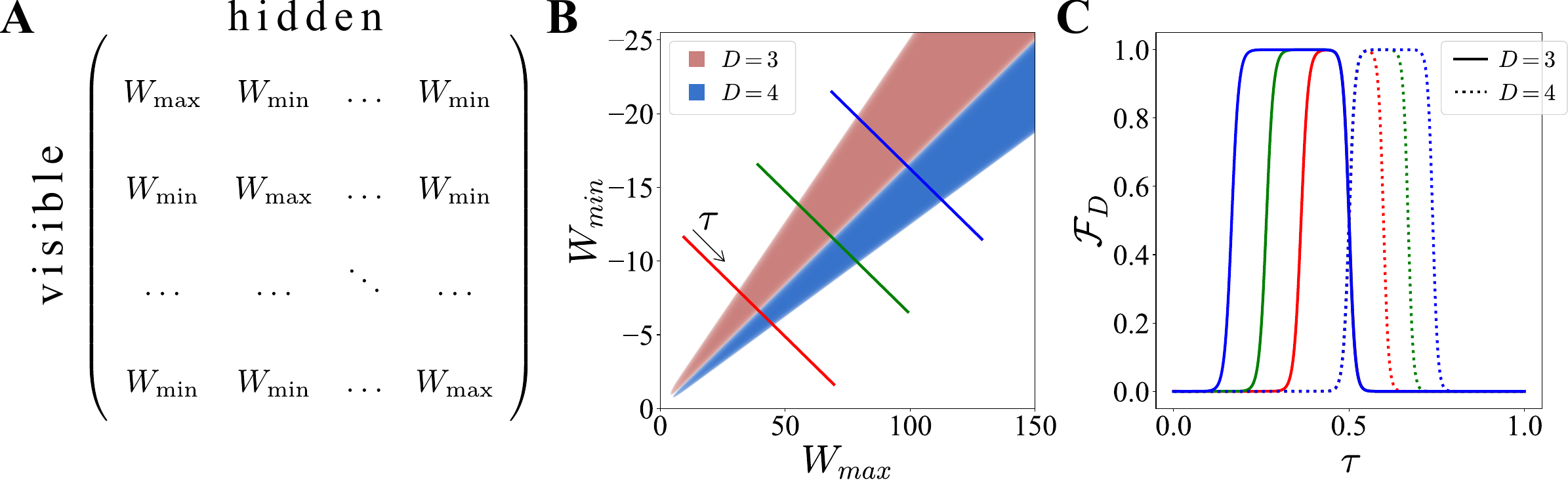}
	\caption{\label{8-qubit} Structural properties and phase diagram of the compact RBM we used to describe the 8-qubit Dicke family. (A) Weight matrix we used to simulate the Dicke states family with RBM. (B) Dicke states sectors in $W_{\text{min}}$-$W_{\text{max}}$ parameter space with $D = 3$ and $D = 4$ for NQS with $N = 8$ visible neurons. Lines denote fidelity calculation paths. (C) Fidelity as function of $\boldsymbol{\lambda}$ defined along three different paths. Lines colors in (B) correspond to line colors in (C).}
\end{figure*}

Previously, the real-valued Dicke states with $D = 1$ (known as the $W$ state) for the systems of 20, 40 and 80 qubits were reconstructed with the neural quantum state tomography \cite{NNtomography}. Since the $\Ket{\Psi^{1}_{N}}$ is the simplest wave function in the Dicke family and is characterized by the smallest entanglement, the RBM used for the tomography contains the same number of visible and hidden units. The reported overlap between the ideal $W$ state and its RBM approximation for $N = 20$ qubits is very high, 0.997. To our knowledge, the tomographic results for the other states with larger $D$ that belong to the family have not been demonstrated. At the same time by increasing the Dicke index $D$ from 1 to $\frac{N}{2}$, one increases the complexity of the quantum state. In our work we focus on these highly-entangled states.   

Fig.~\ref{16_tomography} visualizes the neural quantum state tomography results which are the elements of the weight matrices, $W_{ij}$ between visible and hidden neurons of the RBMs we used to approximate different Dicke states for 8-, 16- and 20-qubit systems. The resulting weight matrices are square, since in all the cases the number of hidden and visible neurons is the same. One can see that for the simplest wave functions with $D=1$ the hidden units show random-like patterns of interactions with neurons of the visible layer. In this case, the absolute values of the $W$ matrix elements do not exceed 9. In agreement with the results of Ref. \onlinecite{NNtomography} the target states with $D=1$ are reconstructed with a high fidelity of 0.99 independent of the system size.   

Remarkably, the solutions of the tomographic problem obtained for more entangled wave functions with $D > 1$ reveal distinct patterns of the elements of the weight matrices. Each hidden unit is characterized by the receptive fields having non-zero weights with all the input neurons. There is a maximal (positive) or minimal (negative) interaction with the specific visible neuron, which indicates weight sharing and is an important feature of convolution \cite{CNN-LeCun}. Weak non-zero couplings to all other neurons of visible layer (Fig.\ref{16_tomography} D) reflect a distinct type of correlations of the input data. Thus, in the case of the RBMs approximating the Dicke states with large $D$, the whole weight space is uniformly covered with such global RFs.  It is worth mentioning that revealing the global RF is possible using a RBM with only a few hidden neurons. In Appendix~\ref{apx:state-tomography} we show the corresponding examples in the case of the 16-qubit wave functions.

The results presented in this section were obtained for 8-, 16- and 20-qubit systems by using the optimization procedure as described in Ref.~\cite{NNtomography} with the same number of input and hidden neurons. As one would expect, the increase of the Dicke index $D$ leads to a considerable degradation of reconstructing the quantum states (Fig.~\ref{16_tomography}), the corresponding fidelities between exact wave functions and their RBM approximations decrease to 0.6 in the case of $\Ket{\Psi^{10}_{20}}$. To improve the agreement the number of the hidden units should be substantially increased so that the overlap is close to 90\% (Appendix B). Below we show that when constructing RBM to approximate a Dicke wave function one can work around the hard problem of the neural quantum tomography by utilizing the architecture constructed with the revealed global receptive fields. 

\begin{figure*}[!t]
	\includegraphics[width=\linewidth]{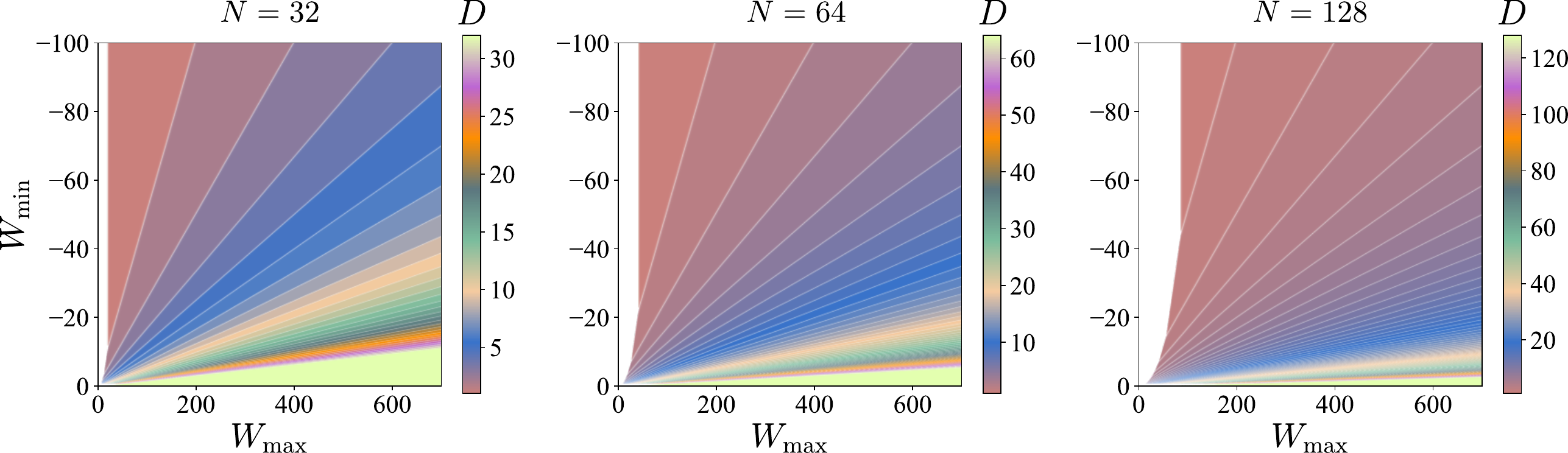}
	\caption{\label{fig:dicke_bruteforce_many} Dicke states diagram for different system sizes obtained with classical RBM representation developed in this work. Colorbars denote value of the $D$ parameter and white areas indicate superposition of different Dicke states such that fidelity of NQS with any of the Dicke states is not larger than 0.5.}
\end{figure*}

\subsection{A compact and exact RBM representation for all $N$-qubit Dicke states with global RFs}
As shown in Ref.\cite{NNtomography} the $\Ket{\Psi^{D}_{N}}$ wave function can be represented exactly with RBM, however, for that one should reserve single hidden unit for each pair of qubits in the system in question. This means the total number of parameters grows as $N^2$ as the system size increases. The most compact representation of a $N$-qubit Dicke state with arbitrary $D$ in terms of the number of hidden neurons so far has been reported in Ref.\cite{Clark1}. In this work the author has used a correlator product state approach to encode $\Ket{\Psi^D_N}$ and analytically showed that any Dicke wave function can be described with $M = [\frac{N}{2}]$. Here, we are going to demonstrate that by using global receptive fields one can develop exact and compact RBM representation for the whole Dicke wave functions family with $M=N$.   

To imitate the Dicke states with RBM, we employ a non-sparse circular square weight matrix, $W$ that is schematically presented in Fig.\ref{8-qubit} and is filled with only two real-valued parameters, $W_{\rm min}$ and $W_{\rm max}$. Taking into account the structure of the discovered global receptive field, in our model RBM each hidden neuron has a positive $W_{\rm max}$ interaction with one unique visible unit and negative $W_{\rm min}$ with other visible sites. This means that the permutation symmetry, basic property of the Dicke states is violated at the level of an individual hidden neuron. To restore the permutation invariance of the whole two-parameter RBM representation the total number of hidden neurons is taken to be equal to $N$ and their connectivity is synchronized with respect to the $W_{\rm max}$ couplings. Thus, $M=N$, $W$ is the square matrix and according to Ref.\cite{Clark2} our RBM can be classified as compact one. The biases for visible and hidden neurons are set to zero, $\vec{a}=\vec{b}=0$. Constraining weight matrix structure in such a way makes the amplitudes of the basis states with the same number of ''1''s to be the same. This ensures that the RBM-based wave function respects the permutation symmetry, which is a fundamental property of the Dicke states.

All the basis functions can be distinguished by the parameter $d = \sum_{i} v_{i}$. For the given $W_{\rm min}$ and $W_{\rm max}$ the basis functions, $\Ket{\bf v}$ that are characterized by the same $d$ have the same amplitude 
\begin{eqnarray}
\label{eq:rbm:Dicke}
\Psi_{\boldsymbol{\lambda}}(d) = \frac{(1+ e^{W_{\rm max} + (d-1) W_{\rm min}})^{\frac{d}{2}} (1+ e^{d W_{\rm min}})^{\frac{N-d}{2}}}{\sqrt{Z_{\boldsymbol{\lambda}}}}. 
\end{eqnarray}
Thus, the RBM wave function of the system in question can be rewritten as
\begin{equation}
  \ket{\Psi_{\boldsymbol{\lambda}}} = \sum_{\bf v}  \Psi_{\boldsymbol{\lambda}} \left(d[{\bf v}]\right) \Ket{\bf v},
\end{equation}
where $d[{\bf v}] = \sum_{i} v_{i}$.

To construct the $W_{\rm min}$ - $W_{\rm max}$ phase diagram that distinguishes the RBM states with respect to the Dicke wave functions with different indices $D$, the continuous parameter space is first discretized with a mesh. Then, for each point of the mesh the fidelities, $\mathcal{F}_D$ between the neural quantum state $\Ket{\Psi_{\boldsymbol{\lambda}}}$ and all the Dicke states are calculated. More specifically, one defines a set of fidelities $\mathcal{F}_D = |\braket{\Psi^D_N|\Psi_{\boldsymbol{\lambda}}}|^2$, where $0 \le D \le N $.
In general case, such fidelity calculations require determination of the normalization constant, $Z_{\boldsymbol{\lambda}}$ and summation over the whole basis states, which cannot be performed for large number of qubits $N$. Fortunately, the proposed architecture of the neural quantum state allows to directly estimate its fidelity with any Dicke state by means of the following expression (Appendix D):
\begin{equation}
  \label{eq:exact_fidelity}
  \mathcal{F}_D = \left(1+ \sum_{d\neq{}D}\frac{\tilde{p}_{\boldsymbol{\lambda}}(d)}{\tilde{p}_{\boldsymbol{\lambda}}(D)}\frac{(N-D)!D!}{(N-d)!d!}\right)^{-1}
\end{equation}
Here, $\tilde{p}_{\boldsymbol{\lambda}}(d)$ are unnormalized RBM probabilities (Eq.\ref{eq:rbm} without normalization constant) for basis functions with Dicke index $d$. In this way one can attribute a given point in weight parameter space $\boldsymbol{\lambda} = \left\{W_{\rm min},W_{\rm max}\right\}$ to specific Dicke state or a composition of Dicke states. 

Fig.~\ref{8-qubit} B gives an example where the neural networks with parameters denoted with red and blue colors are characterized by the 100\% fidelity with the $D=3$ and $D=4$ Dicke states, respectively. Importantly, the narrow transition area (Fig.~\ref{8-qubit} C) between two sectors corresponds to a superposition of the aforementioned Dicke wave functions. To demonstrate the feasibility of our approach for representing entangled quantum states with global receptive fields in Fig.~\ref{fig:dicke_bruteforce_many} we show the complete phase diagrams of the Dicke states for the systems with $N$ = 32, 64 and 128 qubits. 

It is also possible to derive a condition on $W_{\max}>0$ and $W_{\min}<0$ that maximizes the fidelity for a given Dicke state. For that one takes the RBM energy for the particular configuration of the visible ($\bf v$) and hidden ($\bf h$) units as $E_\lambda ({\bf v}, {\bf h}) = - \sum_{ij} W_{ij} v_{i} h_{j}$ and the corresponding probability has the form $p_{\lambda} ({\bf v}, {\bf h}) = \frac{1}{Z_{\lambda}} e^{-E_\lambda ({\bf v}, {\bf h})}$.
Then, in the case when $K$ of $N$ visible neurons are active, we are to find the optimal configuration of hidden neurons that minimizes the RBM energy. Enabling any hidden neuron that is not connected to $K$ active visible neurons through the $W_{\rm max}$ coupling increases the RBM energy by $KW_{\min}$, so these neurons are preferably inactive.
At the same time, enabling $K$ remaining hidden neurons causes a shift in energy of $\Delta E = - W_{\max} - (K-1) W_{\min}$. 
The energy will be minimized by activating all $K$ hidden neurons connected to $K$ activated visible neurons through the $W_{\rm max}$ coupling. 
The RBM energy in this case is $K\Delta E = - KW_{\max} - K(K-1)W_{\min}$. The minimum over $K$ is obtained when $K = (1 - W_{\max}/W_{\min})/2$. 
In this way, unit fidelity for $\ket{\Psi_N^D}$ is achieved when 
\begin{equation}
    \frac{W_{\max}}{-W_{\min}} = 2D-1
\end{equation}
and $W_{\max}$ approaching infinity, which suppresses finite-temperature fluctuations. We observe that the RF-type symmetrical structure of the RBM weights ``selects'' states in the visible (and also the hidden) layers with a specific number of excitations. The high nonlocality of the weight matrix, shown in Fig.~\ref{8-qubit}, manifests itself through a complex pattern of correlations among the resulting Dicke states, which is demonstrated in Fig.~\ref{Ursell}.

As soon as the RBM parameters are attributed to the particular Dicke state the corresponding entanglement entropy can be calculated analytically \cite{Dicke_entropy} for any bipartition. In the same spirit one can define genuine multipartite correlations for $\Ket{\Psi^D_N}$ \cite{Multipartite_correlations}.

\subsection{Comparison with known classical representations}
We have demonstrated that by using global receptive fields one can develop exact and compact RBM representation for the whole Dicke wave functions family with $M=N$. Since the Dicke states attract considerable attention in the fields of condensed matter physics and quantum computing, it is important to compare the constructed RBM-based classical representation with existing analogs. 
As shown in Ref.~\cite{NNtomography}, the $\Ket{\Psi^{D}_{N}}$ wave function can be represented exactly with RBM, however, for that one should reserve single hidden unit for each pair of qubits in the system in question. This means the total number of parameters grows as $\frac{N(N-1)}{2}$ as the system size increases, and evidences that the RBM representation of the Dicke states we propose is more efficient. 

To date the most compact classical representation of a $N$-qubit Dicke state with arbitrary $D$ so far has been reported in Ref.~\cite{Clark1} where the author has used a correlator product state (CPS) approach~\cite{CPS} to encode $\Ket{\Psi^D_N}$. More specifically, in the CPS representation, the amplitude of a basis state of the $N$-qubit Dicke wave function with even $D$ is given by the product of the following correlators,\cite{Clark1}
\begin{eqnarray}
\label{CPS}
\Psi_{\rm CPS} ({\bf v}) = \Upsilon^{\rm triv}_{\bf v}  \Upsilon^{\rm odd}_{\bf v}  \prod_{m({\rm even}), m \neq D}   \Upsilon^{m \Rightarrow 0}_{\bf v}.
\end{eqnarray}
Here the correlators $\Upsilon^{\rm triv}_{\bf v}$ and  $\Upsilon^{\rm odd}_{\bf v}$ cancel the trivial basis state ($\ket{000...0}$ and $\ket{111...1}$) and the basis states with odd number of ``1''s, respectively. In turn, the correlators $ \Upsilon^{m \Rightarrow 0}_{\bf v}$ are aimed to cancel out the basis states with even number of  the ``1''s except those corresponding to the sector with $D$. There is an analogous combination of correlators for the case of the odd $D$~\cite{Clark1}. The total number of the correlators required to encode any Dicke wave function can be estimated as $[\frac{N}{2}]$, where $[.]$ denotes taking integer part. This means that $N[\frac{N}{2}]$ correlators are needed to describe the whole Dicke family.

Since RBM and CPS representations of quantum wave functions are closely related to each other \cite{Clark1}, it is instructive to compare them by the example of the Dicke states we analyze. Each correlator in the CPS representation of $\Ket{\Psi^{D}_{N}}$ (Eq.\ref{CPS}) has a distinct structure and is aimed at filtering a specific part of the Hilbert space. In turn, hidden units in our RBM representation are of the same structure, which results in equilibration of the amplitudes of the basis states corresponding to the same number sector with a definite number of "1"s (Eq.\ref{eq:rbm:Dicke}). The balance between probabilities of different sectors in the RBM representation can be tuned with two parameters. As we have shown above  one can switch between different Dicke state by changing the values of  $W_{\rm min}$ and $W_{\rm max}$.
 
Still another one approach that enables a classical representation of the Dicke states is a matrix product state (MPS) \cite{MPS1, MPS2}. In this method one introduces auxiliary degrees of freedom to generate correlations between physical spins (qubits) and each amplitude is represented as a product of $\chi \times \chi$ matrices, where the bond dimension parameter $\chi$ depends on the entanglement of the target quantum state. As it was shown in Ref.~\cite{MPS1}, the simplest $N$-qubit Dicke state with $D$=1 can be defined using MPSs with periodic boundary conditions and $\chi = 2$. The corresponding classical representation of the wave function is given by 
\begin{eqnarray}
\Ket{\Psi_{\rm MPS}} = \sum_{v_1...v_N = 0}^{1} {\rm Tr}(A^{[1]}_{v_1} A^{[2]}_{v_2}...A^{[N]}_{v_N})\Ket{v_1 ... v_N}.
\end{eqnarray} 
Here $v_i \in \{0,1\}$ denotes the states of the individual qubits and $A^{[i]}_{v_i}$ is the $2 \times 2$ matrix corresponding to the basis state $\ket{v_{i}}$ of the $i$th qubit. $A^{[i]}_{v_i}$ can be defined through  the Pauli matricies as $\{ A^{[i]}_{0}, A^{[i]}_{1} \} = \{ \mathbb{1}, \sigma^{+} \}$ for $i < N$ and $\{ A^{[i]}_{0}, A^{[i]}_{1} \} = \{ \sigma^{x}, \sigma^{+} \sigma^{x} \}$ for $i = N$. Despite reproducing the correct amplitude structure, by the construction such a representation with site-dependent $A^{[i]}_{v_i}$ does not respect the translation symmetry of the Dicke states. The translationally-invariant (TI) form of the MPS for $\Ket{\Psi^{1}_{N}}$ has been derived in Ref.~\cite{MPS3}, where the authors have used the matrices $A_{v_i}$ of size $[\frac{N}{2}]+1$. According to Ref.~\cite{MPS4}, an arbitrary Dicke state $\Ket{\Psi^{D}_{N}}$ can be defined using non-TI MPSs with open boundary conditions and minimal bond dimension $\chi = D+1$. Since by the construction a MPS follows a chain geometry,  searching for a suitable ordering of the MPS variables when describing a quantum state is still a vast open problem~\cite{MPS5}. 

Absence of the direct couplings between visible neurons in the RBM representation generally provides a flexibility in respecting different symmetries of underlying quantum states. For instance, our RBM formulation for the Dicke states, Eq.~\eqref{eq:rbm:Dicke} is permutationally invariant.   

\begin{figure*}[!t]
	\includegraphics[width=\linewidth]{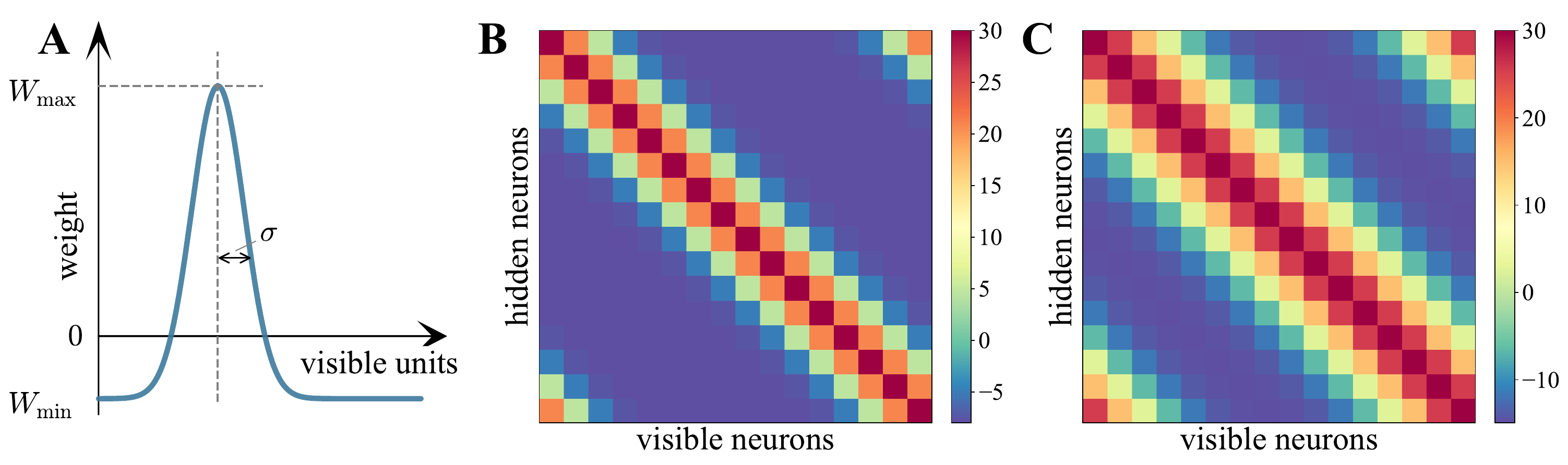}
	\caption{\label{fig:receptive_gaussian} (a) Schematic representation of the global receptive field characterized by a Gaussian smearing, that is controlled with the parameter $\sigma$. (b) and (c) show two possible choices of the RBM weight matrices encoding the quantum state $\ket{{\rm T}^{8}_{16}}$. The corresponding RBM parameters are $W_{\rm min}$ = -8, $W_{\rm max}$ = 30, $\sigma$ = 1.35 for (b) and $W_{\rm min}$ = -15, $W_{\rm max}$ = 30, $\sigma$ = 2.2 for (c). }
\end{figure*}

\subsection{Quantum state design with receptive fields}
Having described the global receptive fields that enable the accurate and concise neural network representation of Dicke wavefunctions, we are now in a position to demonstrate how to develop analogous classical representations for other specific quantum states using our approach.
For these purposes, we start with the RBM weight matrix schematically represented in Fig.~\ref{8-qubit} and modify it by introducing a finite Gaussian smearing of the initial single-neuron peaks for each hidden unit. The resulting receptive field profile [Fig.\ref{fig:receptive_gaussian}(a)] is characterized by a finite-size standard deviation, $\sigma$ that becomes an additional parameter of the RBM wave function. Similarly to the case of the Dicke states that corresponds to $\sigma = 0$, hidden neurons share the same global RF function of couplings with visible layer neurons and the resulting weight matrix is circular [Figs.~\ref{fig:receptive_gaussian}(b) and (c)].

Varying the parameters $W_{\rm min}$, $W_{\rm max}$ and $\sigma$ at the zero biases (${\bf a} = {\bf b} =0$) one can generate a distinct family of  translationally invariant $N$-qubit quantum states
\begin{eqnarray}
\Ket{{\rm T}^D_N} = \frac{1}{\sqrt{N}} \sum_{j} \mathcal{T}_{j}(\Ket{0}^{\otimes N-D} \otimes \Ket{1}^{\otimes D}),
\label{T_wf}
\end{eqnarray}
where the sum runs over $N$ translations of the single block of $D$ excitations ("1"s). Each basis function contributing to $\Ket{{\rm T}^D_N}$ is characterized by the sequence of the $D$ qubits with the state $\ket{1}$ taking into account periodic boundary conditions.
A 6-qubit example of these translationally invariant quantum states with three excitations that can be obtained with Gaussian-like global receptive fields is given by
\begin{eqnarray}
\Ket{{\rm T}^{3}_{6}} = \frac{1}{\sqrt{6}}(\ket{000111}+ \ket{001110} + \ket{011100}+ \nonumber \\  \ket{111000}+\ket{110001}+\ket{100011}).
\end{eqnarray}

\begin{figure*}[!t]
	\includegraphics[width=\linewidth]{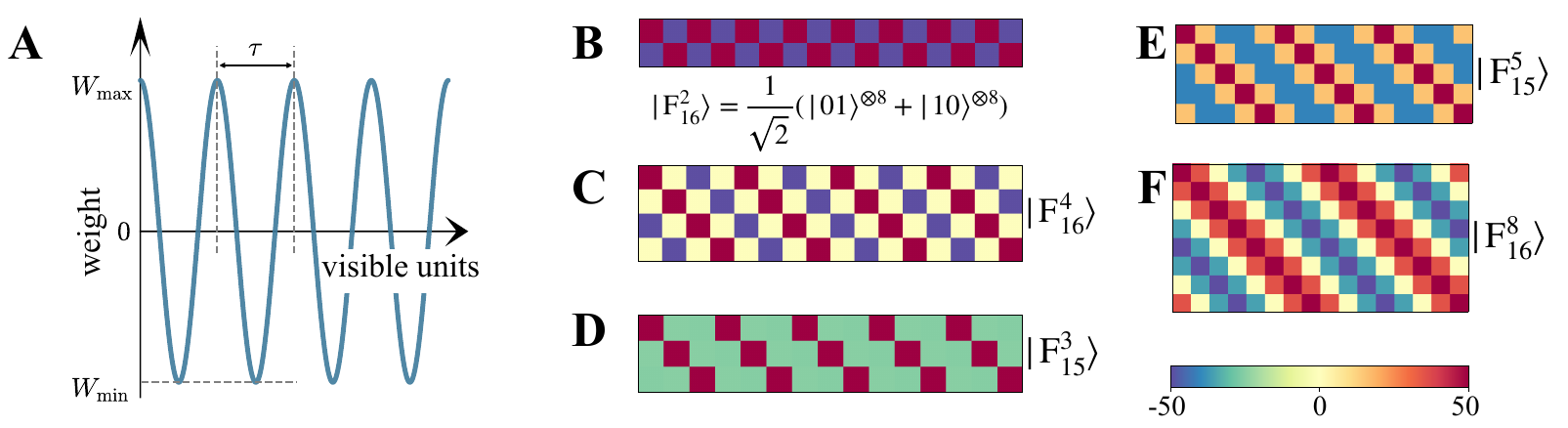}
	\caption{\label{fig:receptive_cosine} (A) Schematic representation of the global receptive field with cosine profile for the given hidden neuron. The period of the oscillations in the weight space is denoted as $\tau$. (B) - (F) Visualizations of the $M\times N$ matrices of the RBM weights between $M$ hidden neurons and $N$ visible units. Each pattern corresponds to the specific wave function $\ket{F^{\tau}_{N}}$ that is represented with a RBM neural network. For these wave functions we use $W_{\rm max} = -W_{\rm min} = 50$.}
\end{figure*}

For $D = 1$ the introduced quantum states coincide with the simplest Dicke wave functions, the $W$ states. $\Ket{{\rm T}^D_N}$ differs from $\Ket{\Psi^{D}_{N}}$ for $D \in [2, N-2]$ and in this regime such periodic wave functions can be considered as solution of a model. These wave functions can be differentiated at the level of the parent Hamiltonians for which $\Ket{{\rm T}^D_N}$ and $\Ket{\Psi^{D}_{N}}$ are the ground states.  For instance, it was shown that the Dicke states are the ground wave functions of the Lipkin-Meshkov-Glick (LMG) model with ferromagnetic couplings defined on the complete graph \cite{LMG1, LMG2, LMG3},
\begin{eqnarray}
\hat H_{\rm LMG} = - \sum_{i < j } (\hat \sigma^x_{i} \hat \sigma^x_{j} + \hat \sigma^y_{i} \hat \sigma^y_{j}) - h \sum_{i} \hat \sigma^z_{i},
\end{eqnarray}
where $h$ is the external magnetic field and $\sigma_i^{\mu}$ is a Pauli matrix. Varying the value of $h$ one can switch between different Dicke states \cite{Dicke_sign}  

In turn, we have found that $\Ket{{\rm T}^D_N}$ is the ground state for the Hamiltonian of the form
\begin{equation}\label{eq:ring-hmlt}
    \hat H = (\hat n_{\rm tot} - D)^2 - \epsilon \sum_{i=1}^N\hat\sigma^z_i\hat\sigma^z_{i+1},
\end{equation}
where $\hat n_{\rm tot}=\sum_{i=1}^N(1-\hat\sigma_i^z)/2$ is the ``excitation number'' operator, $\sigma_i^z$ is the standard $z-$Pauli operator for $i$-th qubit, and $0<\epsilon < N^{-1}$ promotes an appearance of a single domain of excitations within the circular chain (here we set $n+1\equiv1$).
Note that the Hamiltonian~\eqref{eq:ring-hmlt} includes all-to-all couplings between $N$ qubits due to the first term $(\hat{n}_{\rm tot}-D)^2$.

To demonstrate flexibility of our approach we also exploit a completely different shape of the global receptive field that is described with a cosine function presented in Fig.~\ref{fig:receptive_cosine}. The period of the global receptive field, $\tau$ in the space of a single hidden neuron controls the oscillations of the excitations on the level of the basis functions of the resulting quantum state. Similar to the cases of $\ket{\Psi^D_N}$ and $\ket{{\rm T}^{D}_{N}}$ wave functions the $M$ hidden neurons with cosine-like receptive fields should uniformly cover the space of the $N$ visible neurons, which can be done with $M$ smaller than $N$. For instance, setting period of oscillations $\tau=2$ for the system with even number of qubits we obtain $\ket{{\rm F}^{2}_{N}} = \frac{1}{\sqrt{2}}(\ket{01}^{\otimes \frac{N}{2}}+ \ket{10}^{\otimes \frac{N}{2}})$ with only two hidden neurons. Such an entangled wave function is nothing, but the ground state of the quantum Ising antiferromagnets with nearest neighbour interactions on the one-dimensional chain or two-dimensional square lattices with periodic boundary conditions,
\begin{eqnarray}
H_{\rm Ising} = - \sum_{ij} J_{ij} \hat \sigma^z_{i} \hat \sigma^z_{j}.
\end{eqnarray}

Below we give other examples of the wave functions from the $\ket{{\rm F}^{\tau}_{N}}$ family with odd and even number of qubits, 
\begin{align}
\ket{{\rm F}^{4}_{16}} & =  \frac{1}{2}(\ket{1100}^{\otimes 4}+\ket{0110}^{\otimes 4}+\ket{0011}^{\otimes 4} + \ket{1001}^{\otimes 4}), \\
\ket{{\rm F}^{3}_{15}} &=  \frac{1}{\sqrt{6}}(|100\rangle^{\otimes 5}+|010\rangle^{\otimes 5} + |001\rangle^{\otimes 5} + |110\rangle^{\otimes 5} \nonumber \\ &+ |101\rangle^{\otimes 5} + |011\rangle^{\otimes 5}).
\end{align}
These translationally invariant wave functions have exact RBM representations and the corresponding weight matrices between hidden and visible neurons are visualized in Fig.\ref{fig:receptive_cosine}. The characterization of the quantum correlations in the introduced $\Ket{{\rm T}^D_N}$ and $\ket{{\rm F}^{\tau}_{N}}$ states is left for a future investigation.

\section{Discussions and perspectives}

Deep convolution networks with local receptive fields are known to be computationally inefficient for capturing large-scale correlations in data of different origin, which stimulates developing distinct types of the neural models such as non-local neural networks~\cite{Wang} and transformers~\cite{transformer_original}. Constructing non-local receptive fields to extend the capabilities of CNN and other architectures is problem specific and hard to do from scratch. However, our study devoted to approximation of the entangled Dicke wave functions shows that one can employ additional unsupervised learning tools such as quantum neural state tomography and, in this way, get basic information about optimal structure of the receptive fields within the learning procedure. Based on our results concerning the Dicke states such a learning does not have to be perfect. Additionally, by using the developed concept of the global receptive fields we have introduced two distinct families of the translationally invariant quantum states that are characterized by compact classical representations. 

In this study we fully concentrate on finding classical representations for real-valued positive wave functions, the use of which has a number of both limitations and advantages. For instance, as was shown in Ref.\cite{Grover} the entanglement entropy of a typical random positive wave function do not scale with volume, as is possible in general case with complex-valued quantum states. On the other hand, search for local unitary transformations mapping a wave function into a positive-real form (pozitivization) \cite{Marshall, Torlai} is of a special interest in condensed matter physics, since a nontrivial sign (phase) structure prevents one from simulating interesting phases of matter \cite{Bagrov1, Bagrov2, Titus}. As a promising direction for future investigations, we consider searching for receptive fields to represent complex-valued wave functions. For that one is to perform  measurements in different bases and approximate quantum state using two RBM networks describing its amplitude and phase within neural quantum state tomography procedure, as discussed in Ref.\cite{NNtomography}. 
This allows one to analyze the weight spaces of the trained RBM approximating ground states of quantum Hamiltonians on regular lattices, for instance Ising or Heisenberg Hamiltonian on the square, triangular or other lattices. In some cases neural network models of the aforementioned wave functions are known~\cite{Clark1, Clark2, NNtomography, Duan_representation, Imada_representation, Ivan_Glasser} and we expect to find receptive fields of different shapes and properties, which facilitates developing distinct compact classical representations for these quantum states through learning. 

The development of the exact and compact classical neural network representations of quantum states with receptive field concept can be also perspective for more accurate estimation of different observables with sampling procedure \cite{Torlai_thesis} and derivation of the exact expressions for entanglement entropy \cite{DasSarma}, which eventually may facilitate establishing the connection between classical correlation functions and entanglement measures, which is an actual problem in the quantum information field that considers the entanglement as the physical resource to transfer information and perform complex computations \cite{Corr_concurr, Ent_vs_corr}.    

\section*{Acknowledgements}
The work of OMS, IAI, EOK, AKF and VVM was supported by the Russian Roadmap on Quantum Computing (Contract No. 868-1.3-15/15-2021, October 5, 2021).
The work of EOK and AKF is supported by the Priority 2030 program at the ``MISIS'' University under the project K1-2022-027.
The authors declare that this work has been published as a result of peer-to-peer scientific collaboration between researchers. The provided affiliations represent the actual addresses of the authors in agreement with their digital identifier (ORCID) and cannot be considered as a formal collaboration between Radboud University and the other aforementioned institutions.

\appendix

\section{Permutation structure and correlations of the Dicke states}
\begin{figure}[!t]
    \includegraphics[width=0.98\columnwidth]{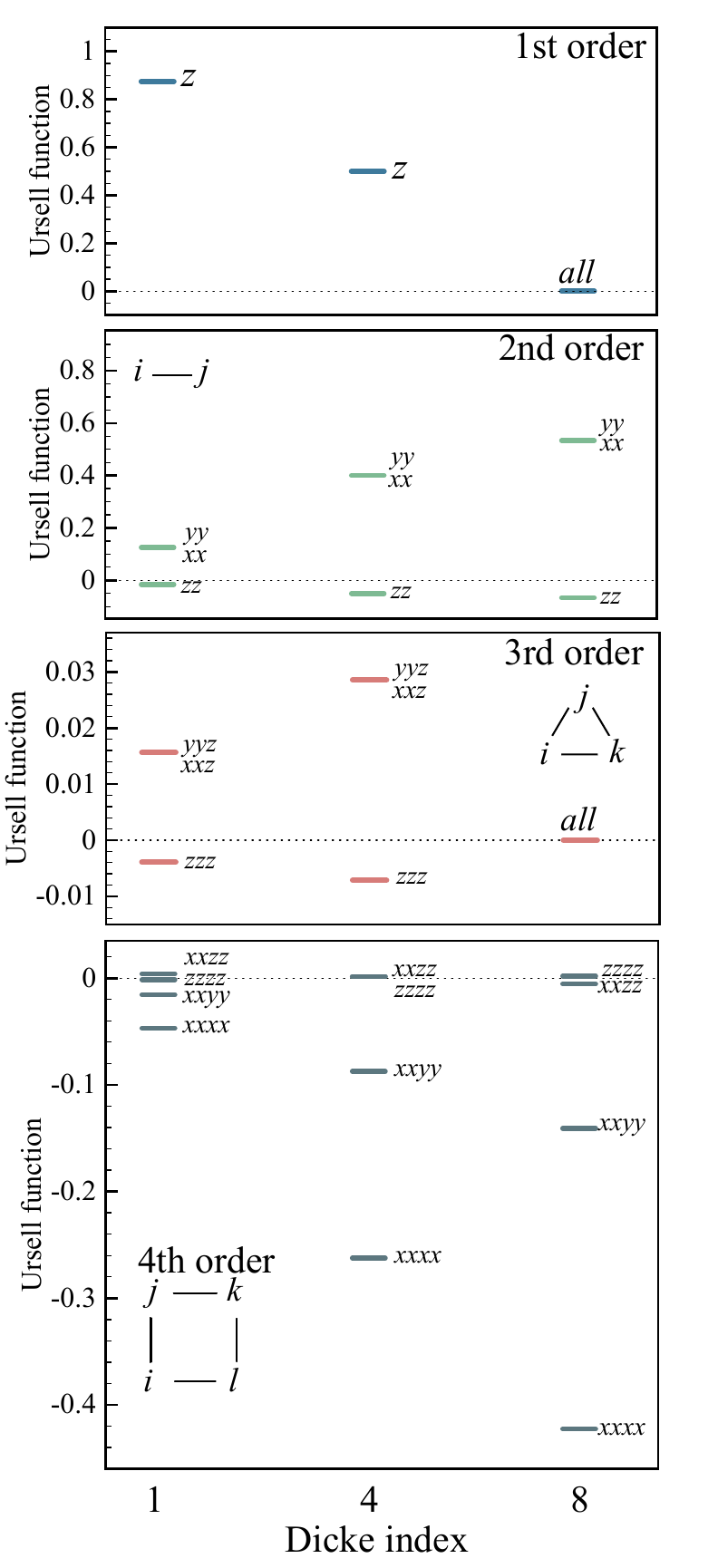}
	\caption{\label{Ursell} Ursell functions up to 4th order that show the correlations structure of the 16-qubit Dicke states. The labels such as $xxzz$ denote spin projection indexes of the calculated correlation functions. In the case of the 3rd and 4th order Ursell functions the correlation levels are multiply degenerate and for presentation purposes each level is associated only with one of correlation functions. The complete correlation histograms are presented in Appendix B. $all$ stands for the cases when all the correlation functions are zero.}
\end{figure}

In this appendix we analyze the properties of the real-valued non-negative wave functions that belong to the Dicke state family,
\begin{eqnarray}
\Ket{\Psi^D_N} = \frac{1}{\sqrt{C^N_D}} \sum_{j} P_{j}(\Ket{0}^{\otimes N-D} \otimes \Ket{1}^{\otimes D}),
\label{Dicke_wf_app}
\end{eqnarray}
where $N$ is the number of qubits, $D$ is the parameter that controls the number of ``1''s in each basis function contributing to the particular quantum state and the sum goes over all possible permutations of qubits, denoted by $P_j$. As a simple example one can consider $N=4$ for which the wave functions with $D=1$ and $D=2$ are given by  
\begin{eqnarray}
	\frac{\Ket{0001} + \Ket{0010} + \Ket{0100} + \Ket{1000}}{2}, \nonumber 
\end{eqnarray}
and 
\begin{eqnarray}
	\frac{\Ket{0011} + \Ket{0110} + \Ket{1100} + \Ket{1001} + \Ket{1010} + \Ket{0101}}{\sqrt{6}}, \nonumber 
\end{eqnarray}
respectively. The former wave function has a simple structure, each contributing basis state can be obtained by translating another one, which corresponds to cyclic permutation of the element 0001 ($0001 \rightarrow 0010 \rightarrow 0100 \rightarrow 1000$).  It is not the case for $\Ket{\Psi^2_4}$ which has a more complicated permutation structure. Here reproducing the whole wave function requires considering two disjoint cyclic permutations for initial elements 0011 and 0101, $0011 \rightarrow 0110 \rightarrow 1100 \rightarrow 1001$ and $0101 \rightarrow 1010$. For larger $N$ the number of such disjoint permutations will increase as the Dicke index, $D$ becomes larger and reach the maximum at $D = \frac{N}{2}$. Thus, from the perspective of constructing a convolutional neural network, exploring Dicke states with $D > 1$ will require a more complicated block of feature maps than that for $D = 1$.  

\begin{figure*}[!t]
    \includegraphics[width=\linewidth]{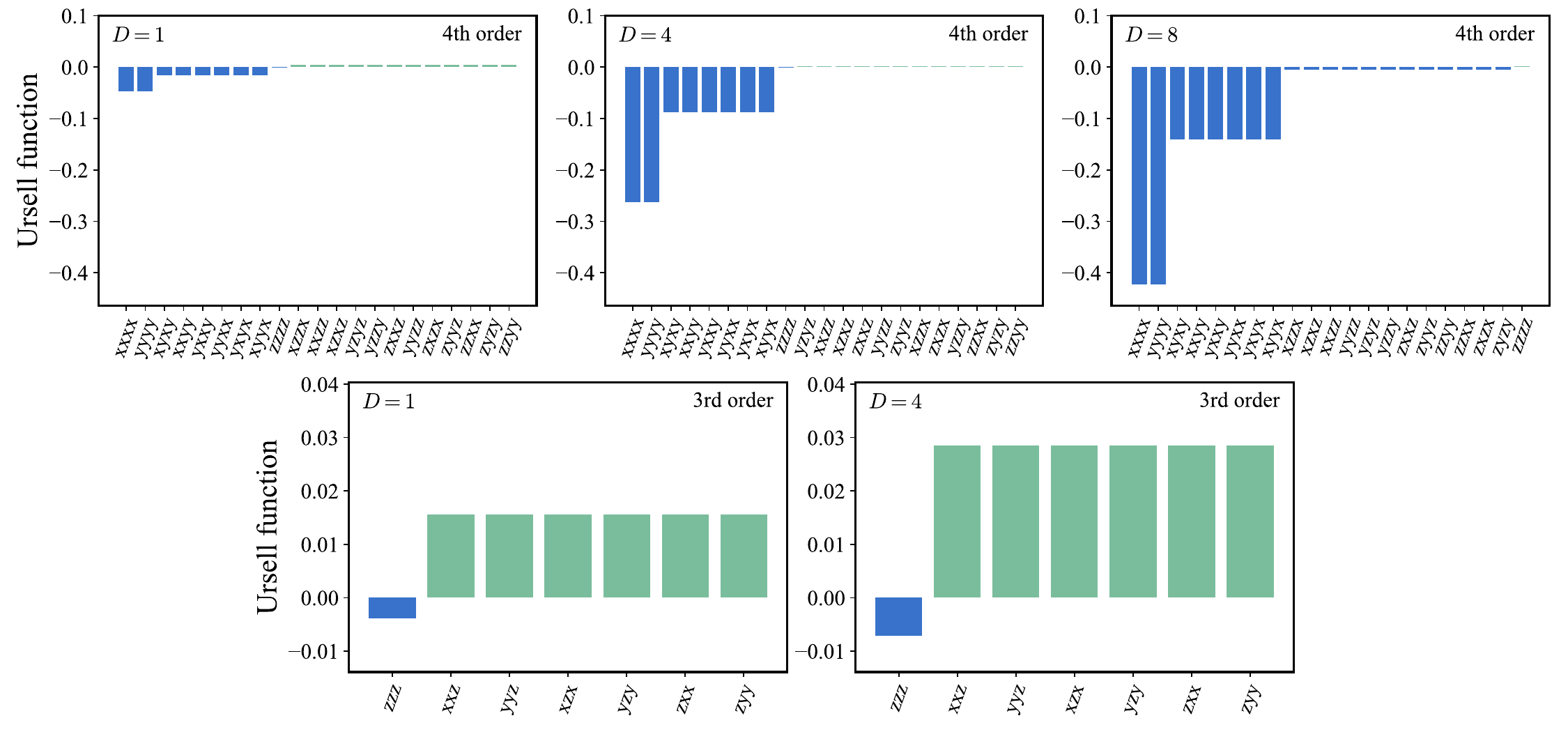}
    \caption{\label{fig:ursell_histograms_all} Non-zero Ursell functions $\Gamma$ for 16-site Dicke states with $D=1,4,8$. The labels such as $xxzz$ denote spin projection indexes of the calculated correlation functions.}
\end{figure*}

The difference in structure of the permutations used for constructing $\Ket{\Psi^D_N}$ with different $D$ parameters is intimately related to variance of quantum correlations of these functions. The previous consideration \cite{Multipartite_correlations} of the quantum correlations on the level of the density matrices with relative entropy has demonstrated that the Dicke states with $D = \frac{N}{2}$ display correlations at any order of a coarse grained partition of the system in question, which evidences a complex multipartite structure of these wave functions. From the perspective of applying a quantum state neural network approach in which one is to approximate unknown probability distribution for the basis functions of a target state, it is instructive to analyze the classical averages that are the correlation functions between qubits. For that we calculate the Ursell functions \cite{Ursell} (connected correlations) up to fourth order for spin operators, ${\hat \sigma}_{i}^{\alpha}$, where $i$ is the site index and $\alpha$ stands for the projection $x$, $y$ or $z$. In our case we concentrate on the first order local correlations, $\Gamma^{\alpha}_{i}$ and non-local correlation functions of higher orders, $\Gamma^{\alpha \beta}_{ij}$, $\Gamma^{\alpha \beta \gamma}_{ijk}$ and $\Gamma^{\alpha \beta \gamma \delta}_{ijkl}$ ($i\ne j \ne k \ne l$). The corresponding expressions for these correlation functions are given in Appendix B.

Fig.~\ref{Ursell} gives the calculated Ursell correlation functions for $D = 1$, 4 and 8 of the 16-qubit system.
In all the cases the connected correlations are invariant with respect to the choice of the qubit indexes and sensitive to the particular combination of the spin projections. In the lowest order we obtain non-zero $\Gamma^{z}_{i}$ only with $D =1$ and 4. In turn, the two-spin $\Gamma$ reveal ferromagnetic correlations that amplify as the Dicke index increases. This can be explained by considering the Dicke wave functions as eigenstates of the Lipkin-Meshkov-Glick model \cite{LMG1,LMG2,LMG3} which is characterized by the ferromagnetic in-plane couplings defined on the complete $N$-node graph. 

As one would expect, increasing the order of the $\Gamma$ function leads to more diverse picture of correlations. At the same time, the values of the non-local $\Gamma^{\alpha \beta \gamma \delta}_{ijkl}$ are not small, as their strength is of the same order of magnitude as those for the two-spin correlation functions. Moreover, increasing the index D clearly demonstrates enhancement of the 4th order $\Gamma$ function, which is a remarkable feature of the Dicke states. This evidences that the correlation structure of $\Ket{\Psi^D_N}$ is substantially non-local and is characterized by high-order excitations. Thus, learning the Dicke states with tomography procedure involves tuning restricted Boltzmann machine in such a way that it can reproduce undamped classical spin correlations of all orders, which would equally mean reconstructing genuine multipartite quantum correlations previously discussed in Ref.~\cite{Multipartite_correlations}.

\section{Ursell functions}

Below, we present the expressions for Ursell functions used to describe the classical correlations of the Dicke states 
\begin{eqnarray}
\Gamma^{\alpha}_{i} = \braket{ {\hat \sigma}^{\alpha}_{i}},
\end{eqnarray}
\begin{eqnarray}
\Gamma^{\alpha \beta}_{ij} = \braket{ {\hat \sigma}^{\alpha}_{i} {\hat \sigma}^{\beta}_{j}} - \braket{{\hat \sigma}^{\alpha}_{i}} \braket{{\hat \sigma}^{\beta}_{j}},
\end{eqnarray}
\begin{eqnarray}
\Gamma^{\alpha \beta \gamma}_{ijk} = \braket{ {\hat \sigma}^{\alpha}_{i} {\hat \sigma}^{\beta}_{j} {\hat \sigma}^{\gamma}_{k}} - \braket{ {\hat \sigma}^{\alpha}_{i}} \braket{{\hat \sigma}^{\beta}_{j} {\hat \sigma}^{\gamma}_{k}} - \braket{ {\hat \sigma}^{\beta}_{j}} \braket{{\hat \sigma}^{\alpha}_{i} {\hat \sigma}^{\gamma}_{k}} -\nonumber \\  - \braket{ {\hat \sigma}^{\gamma}_{k}} \braket{{\hat \sigma}^{\alpha}_{i} {\hat \sigma}^{\beta}_{j}} + 2 \braket{{\hat \sigma}^{\alpha}_{i}} \braket{{\hat \sigma}^{\beta}_{j}} \braket{{\hat \sigma}^{\gamma}_{k}},
\end{eqnarray}
and 
\begin{eqnarray}
\Gamma^{\alpha \beta \gamma \delta}_{ijkl} = \braket{ {\hat \sigma}^{\alpha}_{i} {\hat \sigma}^{\beta}_{j} {\hat \sigma}^{\gamma}_{k} {\hat \sigma}^{\delta}_{l}} - \braket{ {\hat \sigma}^{\alpha}_{i}} \braket{{\hat \sigma}^{\beta}_{j} {\hat \sigma}^{\gamma}_{k} {\hat \sigma}^{\delta}_{l} }  \nonumber \\ - \braket{ {\hat \sigma}^{\beta}_{j}} \braket{{\hat \sigma}^{\alpha}_{i} {\hat \sigma}^{\gamma}_{k} {\hat \sigma}^{\delta}_{l}}  
- \braket{ {\hat \sigma}^{\gamma}_{k}} \braket{{\hat \sigma}^{\alpha}_{i} {\hat \sigma}^{\beta}_{j} {\hat \sigma}^{\delta}_{l}} - \braket{ {\hat \sigma}^{\delta}_{l}} \braket{{\hat \sigma}^{\alpha}_{i} {\hat \sigma}^{\beta}_{j} {\hat \sigma}^{\gamma}_{k}} \nonumber \\ 
-\braket{ {\hat \sigma}^{\alpha}_{i} {\hat \sigma}^{\beta}_{j}} \braket{{\hat \sigma}^{\gamma}_{k} {\hat \sigma}^{\delta}_{l}} - 
\braket{ {\hat \sigma}^{\alpha}_{i} {\hat \sigma}^{\gamma}_{k}} \braket{{\hat \sigma}^{\beta}_{j} {\hat \sigma}^{\delta}_{l}}
-\braket{ {\hat \sigma}^{\alpha}_{i} {\hat \sigma}^{\delta}_{l}} \braket{{\hat \sigma}^{\beta}_{j} {\hat \sigma}^{\gamma}_{k}}
\nonumber \\
+ 2 \braket{ {\hat \sigma}^{\alpha}_{i} {\hat \sigma}^{\beta}_{j}} \braket{{\hat \sigma}^{\gamma}_{k}} \braket{{\hat \sigma}^{\delta}_{l}} + 2 \braket{ {\hat \sigma}^{\alpha}_{i} {\hat \sigma}^{\gamma}_{k}} \braket{{\hat \sigma}^{\beta}_{j}} \braket{{\hat \sigma}^{\delta}_{l}} 
\nonumber \\
+ 2 \braket{ {\hat \sigma}^{\alpha}_{i} {\hat \sigma}^{\delta}_{l}} \braket{{\hat \sigma}^{\beta}_{j}} \braket{{\hat \sigma}^{\gamma}_{k}} + 2 \braket{ {\hat \sigma}^{\beta}_{j} {\hat \sigma}^{\gamma}_{k}} \braket{{\hat \sigma}^{\alpha}_{i}} \braket{{\hat \sigma}^{\delta}_{l}} 
\nonumber \\
+ 2 \braket{ {\hat \sigma}^{\beta}_{j} {\hat \sigma}^{\delta}_{l}} \braket{{\hat \sigma}^{\alpha}_{i}} \braket{{\hat \sigma}^{\gamma}_{k}} + 2 \braket{ {\hat \sigma}^{\gamma}_{k} {\hat \sigma}^{\delta}_{l}} \braket{{\hat \sigma}^{\alpha}_{i}} \braket{{\hat \sigma}^{\beta}_{j}} 
\nonumber \\
 - 6 \braket{{\hat \sigma}^{\alpha}_{i}} \braket{{\hat \sigma}^{\beta}_{j}} \braket{{\hat \sigma}^{\gamma}_{k}} \braket{{\hat \sigma}^{\delta}_{l}}.
\end{eqnarray}
Here $\alpha$ ($\beta$, $\gamma$, $\delta$) stands for spin projection of the Pauli matrix, $\braket{...} = \braket{\Psi^{D}_{N}| ... | \Psi^{D}_{N} }$ denotes the average calculated for the wave function $\Ket{\Psi^{D}_{N}}$. Fig.\ref{fig:ursell_histograms_all} gives the complete histograms of the non-zero Ursell functions of 3rd and 4th orders.

\section{Neural quantum state tomography of the Dicke states} \label{apx:state-tomography}
In this section we describe the details of the neural quantum state tomography whose results are presented in the main text. All the simulations presented in the main text were performed with the following parameters. We used 10 steps within the contrastive divergence procedure and the learning rate was varied in the range  [0.01, 0.2] to reach the best possible fidelity with target state. The number of the training epochs was set to 10000. In some calculations we observed the degradation of the results concerning fidelity with the target quantum state with number of epochs. That is why the neural network weights corresponding to the best fidelity within a run have been selected for further analysis.  

\begin{figure}[!b]
    \includegraphics[width=\linewidth]{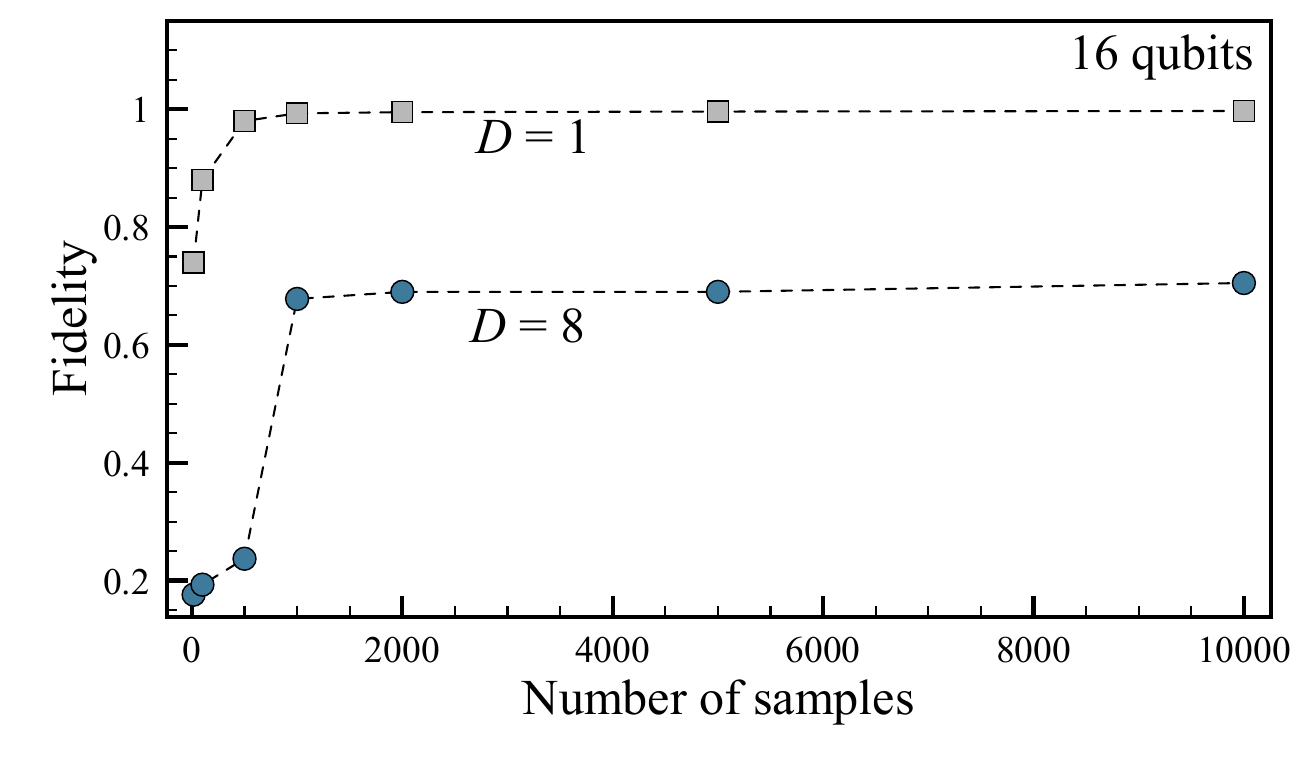}
    \caption{\label{fig:fidelity} Fidelity between the exact 16-qubit Dicke states and their approximations generated by trained RBMs. Grey squares and blue circles denote results obtained for the cases of $D=1$ and $D = 8$, respectively. The number of the hidden units was fixed in these simulations as $M=N=16$.}
\end{figure}

Fig.\ref{fig:fidelity} demonstrates the quality of RBM approximation that has been achieved with different number of the bitstrings in the training set, $N_{s}$ in the case of the 16-qubit Dicke wave functions.  In agreement with Ref.\cite{NNtomography} for $D=1$ ($W$ state) the fidelity approaches to 1 at $10^{3}$ samples, which is due to a strong localization of the target state in the Hilbert space. In the case of the $D=8$ state that is characterized by the largest entanglement for 16-qubit system the saturation value of the fidelity with the exact solution is about 0.7. As follows from Fig.\ref{fig:fidelity}, increasing the number of bitstrings in the training set does not improve the quality of the tomographic procedure.

\begin{figure}
    \includegraphics[width=\linewidth]{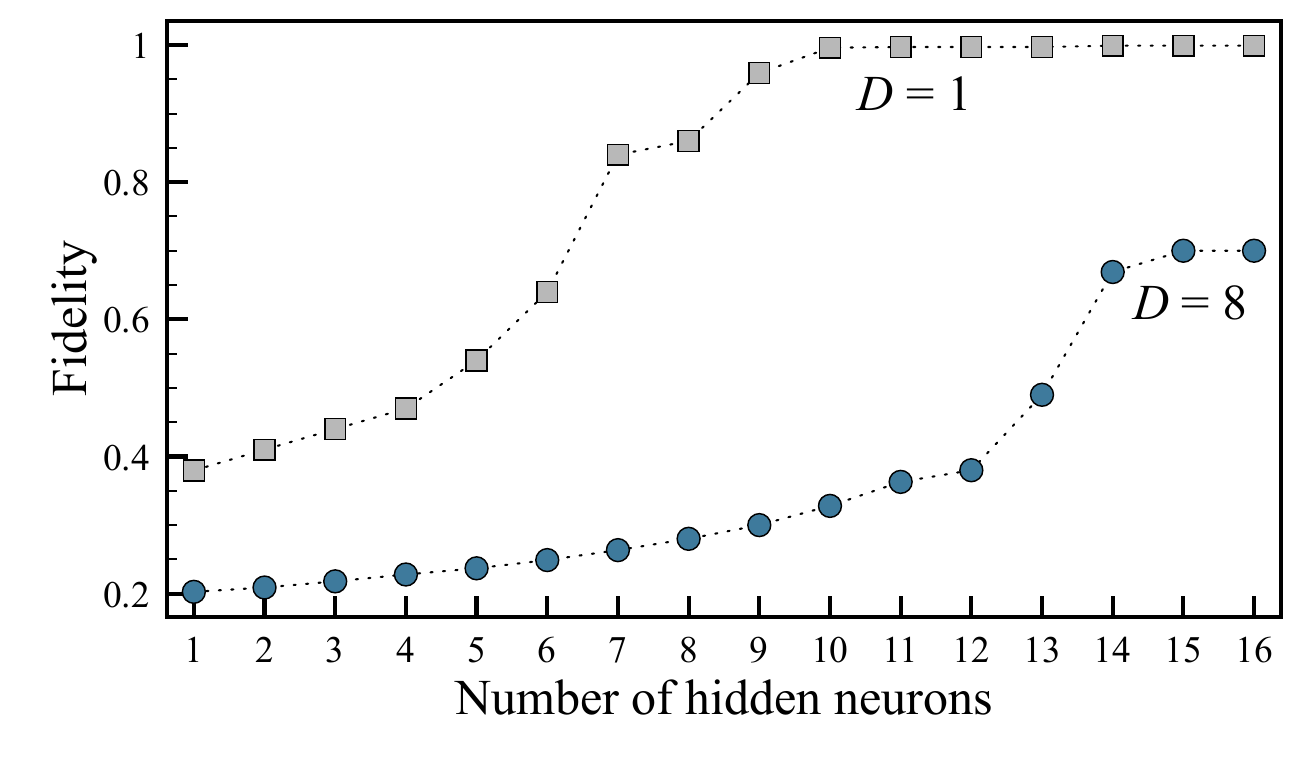}
    \caption{\label{fig:fidelity1-16} Fidelity between ideal 16-qubit Dicke wave functions and their RBM approximations obtained within neural quantum state tomography for different number of the hidden neurons.}
\end{figure}

\begin{figure}
    \includegraphics[width=\linewidth]{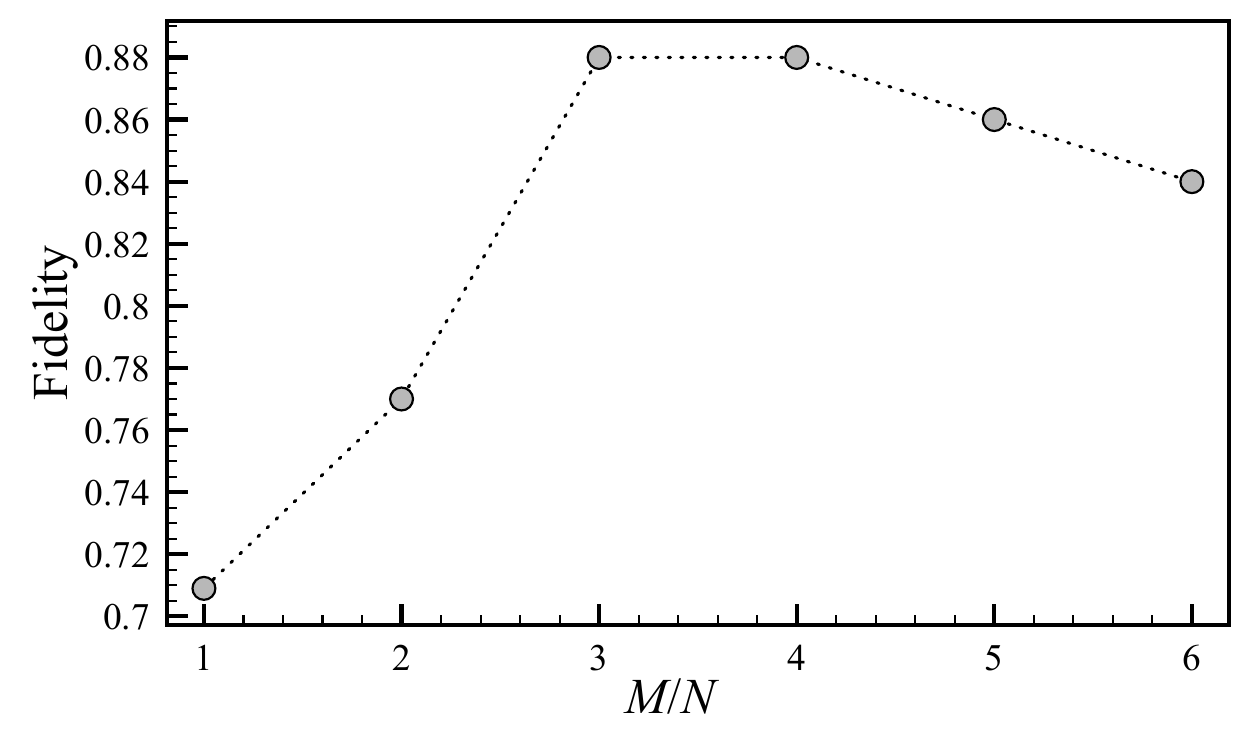}
    \caption{\label{fig:fidelity123456} Fidelity between ideal $\Ket{\Psi^{8}_{16}}$ Dicke wave function and its RBM approximations obtained within neural quantum state tomography for different number of the hidden neurons.}
\end{figure}

The results presented in Fig.\ref{fig:fidelity} and Fig.~3 in the main text were obtained with the same number of the hidden and visible neurons. To complete this consideration it is important to explore the regimes when $M<N$ and $M>N$. 16-qubit examples of the former are given by Fig.\ref{fig:fidelity1-16}. One can see that for $D = 1$ the fidelity of 0.99 with exact solution can be reached with 10 hidden neurons. In turn, quality of the $\ket{\Psi^{8}_{16}}$ approximation is rather low and the corresponding fidelity takes the values of about 0.7 for $M>14$. As described in the main text such a degradation of the fidelity when varying the Dicke index is due to the complexity of the target quantum states that enhances as $D$ increases. This problem can be solved by increasing the number of the hidden neurons (the $M>N$ regime). From Fig.~\ref{fig:fidelity123456} plotted for the $\Ket{\Psi^{8}_{16}}$ state characterized by the largest entanglement it follows that the quality of the neural network approximation can be slightly improved by adding more hidden neurons, which likewise confirms the importance of the RF approach we develop in this work.

\begin{figure}
    \includegraphics[width=\linewidth]{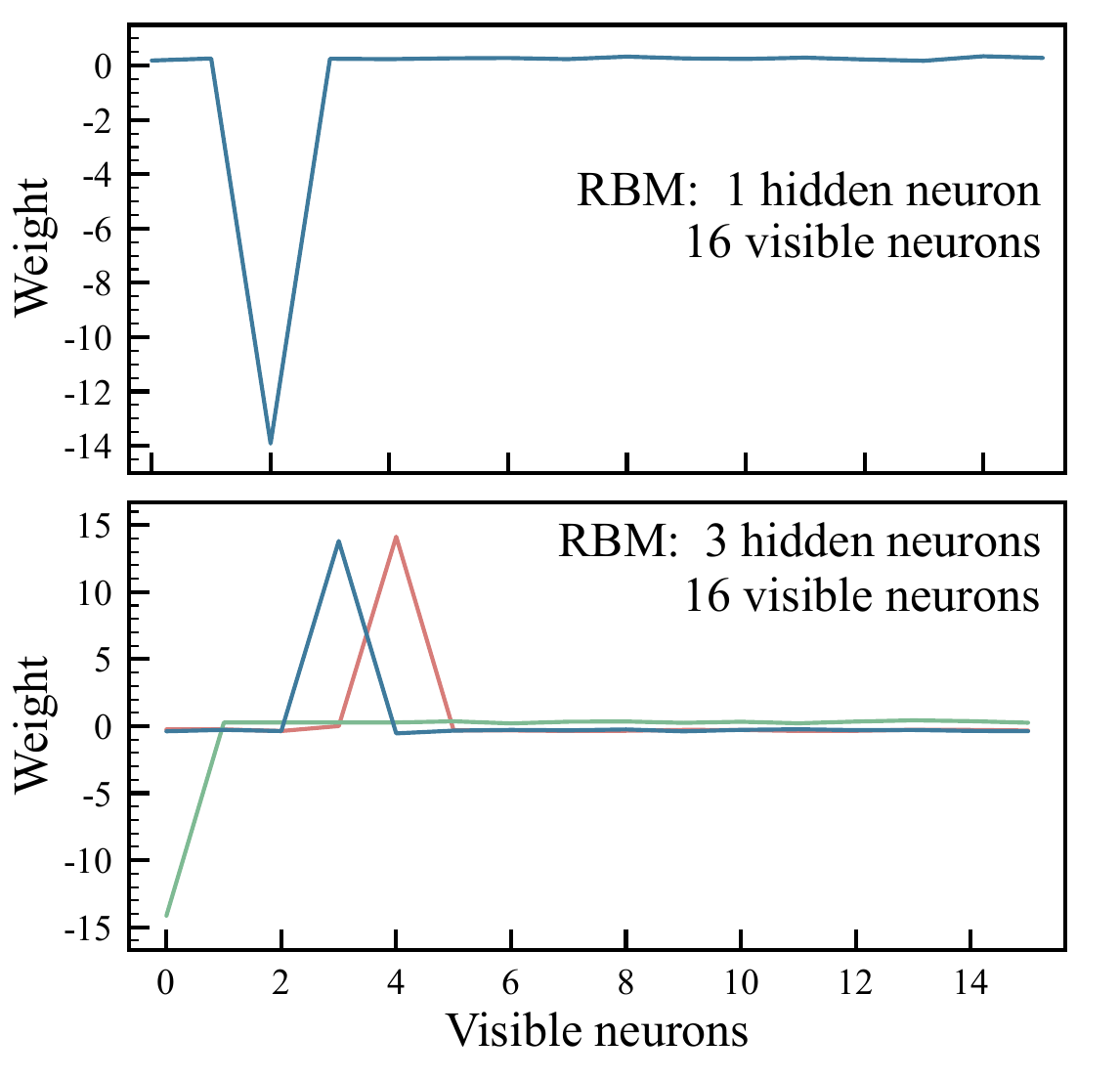}
    \caption{\label{fig:RF_1-3_hneurons} Formation of the global receptive fields in the RBMs with a few hidden neurons. (top) and (bottom) RBM weights between hidden and visible neurons obtained within the neural quantum state tomography of the $\ket{\Psi^{8}_{16}}$. The RBMs feature one (top) and three (bottom) hidden units. The fidelity between RBM approximation and target wave function is 0.20 (top) and 0.22 (bottom).}
\end{figure}

To conclude this section by the example of Fig.\ref{fig:RF_1-3_hneurons} we provide an evidence that RF field can emerge even when one employs a RBM with one or a few hidden neurons. Despite of a low quality of the resulting approximations (in both cases the overlap with the exact state is about 0.2) one can clearly distinguish the profile of the global receptive field formed in the parameter space of the trained RBM with minimal $M$. This leads us to an important conclusion that an optimal network structure could be also developed on the basis of preliminary fast learning with a minimal network setting. 

\section{Derivation of the fidelity expression} \label{apx:fidelity}
To derive Eq.~\eqref{eq:exact_fidelity} in the main text we start with relation between neural quantum state~\eqref{eq:nqs} and Dicke state~\eqref{Dicke_wf}:

\begin{equation}
  \ket{\Psi_{\boldsymbol{\lambda}}} = \sum_{d=0}^N\sqrt{p_{\boldsymbol{\lambda}}(d)C_d^N}\ket{\Psi_N^d}
\end{equation}
where $p_{\boldsymbol{\lambda}}(d)$ are normalized RBM probabilities:
\begin{equation}
  \label{eq:normalized_probabilities}
  p_{\boldsymbol{\lambda}}(d) = Z_{\boldsymbol{\lambda}}^{-1}\tilde{p}_{\boldsymbol{\lambda}}(d) = \frac{\tilde{p}_{\boldsymbol{\lambda}}(d)}{\sum_{i=0}^{N}\tilde{p}_{\boldsymbol{\lambda}}(i)C_i^N}
\end{equation}
Then, one can write the following expression for fidelity:
\begin{equation}
  \label{eq:fidelity_derivation1}
  \mathcal{F}_D = |\braket{\Psi^D_N|\Psi_{\boldsymbol{\lambda}}}|^2 = p_{\boldsymbol{\lambda}}(D)C_D^N
\end{equation}

Substitution of~\eqref{eq:normalized_probabilities} in~\eqref{eq:fidelity_derivation1} gives equation~\eqref{eq:exact_fidelity} in the main text.

\end{document}